# Einstein's gravitational field

**Abstract**: There exists some confusion, as evidenced in the literature, regarding the nature the gravitational field in Einstein's General Theory of Relativity. It is argued here that this confusion is a result of a change in interpretation of the gravitational field. Einstein identified the existence of gravity with the inertial motion of accelerating bodies (i.e. bodies in free-fall) whereas contemporary physicists identify the existence of gravity with space-time curvature (i.e. tidal forces). The interpretation of gravity as a curvature in space-time is an interpretation Einstein did not agree with.

Author: Peter M. Brown  
e-mail: pmb61@hotmail.com

INTRODUCTION

Einstein's General Theory of Relativity (EGR) has been credited as the greatest intellectual achievement of the 20th Century. This accomplishment is reflected in *Time Magazine's* December 31, 1999 issue [1], which declares Einstein the *Person of the Century*. Indeed, Einstein is often taken as the model of genius for his work in relativity. It is widely assumed that, according to Einstein's general theory of relativity, gravitation is a curvature in space-time. There is a well-accepted definition of *space-time curvature*. As stated by Thorne [2]

> *space-time curvature* and tidal gravity are the same thing expressed in different languages, the former in the language of relativity, the later in the language of Newtonian gravity.

However one of the main tenants of general relativity is the *Principle of Equivalence*: *A uniform gravitational field is equivalent to a uniformly accelerating frame of reference*. This implies that one can *create* a uniform gravitational field simply by changing one's frame of reference from an inertial frame of reference to an accelerating frame, which is rather difficult idea to accept. A uniform gravitational field has, by definition, no tidal forces and thus no space-time curvature. Thus according to the interpretation of gravity as a curvature in space-time a uniform gravitational field becomes a contradiction in terms (i.e. no tidal forces where there are tidal forces). This apparent contradiction is obviously quite confusing and can certainly be misleading. In *A brief history of relativity* published in the above mentioned issue of *Time*, Stephen Hawking writes

> I still get two or three letters a week telling me Einstein was wrong. Nevertheless, the theory of relativity is now completely accepted by the scientific community, and its predictions have been verified in countless applications. [...] His idea was that mass and energy would warp space-time in some manner ... Objects like apples or planets would try to move on straight lines through space-time, but their paths would appear bent by a gravitational field because space-time is curved.

Given such a statement by a respected physicists and with a great deal of experimental data to back up this claim is it reasonable to question this notion of gravity being a curvature in space-time? Is it reasonable to assume that the bent path an object takes when moving through a gravitational field is due to the of space-time curvature? Did Einstein actually hold the view that *gravity is a curvature in space-time*? At this point let us whet your appetite. No. Einstein never said nor implied in anyway that *gravity is a curvature in space-time.* This apparent disparity is a result of a change of interpretation. However it would be a great injustice to imply that there is no relationship between gravity and space-time curvature. Curvature plays a very important role in general relativity and its importance should not be underestimated. However the aforementioned change in interpretation is most likely the source of various errors in the scientific literature.

The following is, in part, an historical journey through the origins of general relativity. Our purpose in doing so is to aid in our attempt to certain problematic areas. The following discussion will be address two different interpretations of general relativity. Therefore, for reasons of clarity, Einstein's vision of general relativity, as defined in *The Foundation of the General Theory of Relativity* [3] and other publications, will be referred to as EGR (*Einstein's General Relativity*). General relativity as to it is widely understood at present [2,4,5,6] will be referred to as MGR (*Modern General Relativity*).

In MGR there seems to exists no universally accepted definition of *gravitational field* as noted by Misner, Thorne and Wheeler [7]

> … nowhere has a precise definition of the term "gravitational field" been given --- nor will one be given. Many different mathematical entities are associated with gravitation; the metric, the Riemann curvature tensor, the curvature scalar … Each of these plays an important role in gravitation theory, and none is so much more central than the others that it deserves the name "gravitational field."

The mathematical quantity associated with space-time curvature is the Riemann tensor, also known as the curvature tensor or tidal force tensor. This tensor, as with any tensor, can be expressed as a matrix. It can be shown that if, for a given region of space-time, all of the components of the curvature tensor vanish then that region of space-time is said to be flat. Conversely if a region of space-time is not flat then it's said to be curved.

Curvature is said to be a local quantity in that it is possible to detect its presence in an arbitrarily small region of space in an arbitrarily short amount of time [8]. It is always possible to choose a coordinate system, which is inertial, at least locally [9]. Such a coordinate system is said to be locally flat. Given experimental limitations one can establish criteria in which tidal effects may be ignored [10].

The main goal of this paper is not to present a new interpretation of gravity. For pedagogical purposes we review an old one, that of Albert Einstein's.

HISTORICAL BACKGROUND

This section serves as a brief historical overview of the development of the general theory of relativity and those ideas, which guided Einstein. The main focus of this section will be on the definition of *gravitational field*.

## (1907) On the Relativity Principle and the Conclusions Drawn from It
## Albert Einstein

Following his landmark paper of 1905 Einstein was, for a few years, unable to extend what has come to be known as *special relativity* (i.e. relativity of inertial frames) to non-inertial frames. As Einstein was later to recall [11]

> I was dissatisfied with the special theory of relativity, since the theory was restricted to frames of reference moving with constant velocity relative to each other and could not be applied to the general motion of a reference frame. I struggled to remove this restriction and wanted to formulate the problem in the general case.

Insight towards the solution came late 1907. Einstein, at the request of Johannes Stark, was writing an article for *Jahrbuch der Radioaktivitat und Electronik* [12,13].

> The breakthrough came suddenly one day. I was sitting on a chair in my patent office in Bern. Suddenly the thought struck me: If a man falls freely, he would not feel his own weight. I was taken aback. This simple thought experiment made a deep impression on me. This led me to the theory of gravity. I continued my thought: A falling man is accelerated. Then what he feels and judges is happening in the accelerated frame of reference. I decided to extend the theory of relativity to the reference frame with acceleration. I felt that in doing so I could solve the problem of gravity at the same time. A falling man does not feel his weight because in his reference frame there is a new gravitational field, which cancels the gravitational field due to the Earth. In the accelerated frame of reference, we need a new gravitational field.

This new idea was placed in the fifth section of the Jahrbuch article under *Principle of Relativity and Gravitation* where the following argument was presented

We consider two systems $S_1$ and $S_2$ in relative motion. Let $S_1$ be accelerated in the direction of its X axis, and let *g* be the (temporally constant) magnitude of that acceleration. $S_2$ shall be at rest, but located in a homogeneous gravitational field that imparts to all objects an acceleration -g in the direction of the z-axis.

As far as we know, the physical laws with respect to $S_1$ do not differ from those with respect to $S_2$; this is based on the fact that all bodies are equally accelerated in the gravitational field. At our present state of experience we have thus no reason to assume that the systems $S_1$ and $S_2$ differ from each other in any respect, and in the discussion that follows, we shall therefore assume the complete physical equivalence of a gravitational field and a corresponding acceleration of the reference system.

This assumption extends the principle of relativity to the uniformly accelerated transnational motion of the reference system. The heuristic value of this assumption rests on the fact that it permits the replacement of a homogeneous gravitational field by a uniformly accelerated reference system, the later case being to some extent accessible to theoretical treatment.

As noted earlier this assumption has come to be known as the *Equivalence Principle*. It was here that Einstein first stated that clocks slow down in an accelerating system and gravitational fields. In mathematical terms the relation between the "time" *t* of the system, i.e. the time as reckoned by the clock at the origin of co-ordinates of the accelerated frame, and the "local time" of the system, i.e. the time as reckoned by a clock at *z* in the accelerated system, was found to be

$$= (1 + gz/c^2)t \qquad <\text{Eq. 1}>$$

where c is the speed of light as measured in an inertial frame. This relation was not assumed to be valid everywhere or for all magnitudes of acceleration. *g* was restricted such that terms of the second or higher power may be neglected and *z* restricted as follows

This equation holds first of all if *t* and *z* lie below certain limits. It is obvious that it holds for arbitrarily large     if the acceleration *g* is constant with respect to S, because the relation between     and *t* must be linear. <Eq. 1> does not hold for arbitrarily large *z*. From the fact that the choice of coordinate origin must not affect the relation, one must conclude that, strictly speaking, equation <Eq. 1> should be replaced by

$$ = t\, e^{(gz/c^2)} $$

Nevertheless, we shall maintain formula <Eq. 1>.

If *g* is the acceleration due to gravity on earth then relativity sets an upper bound of approximately 1 light year for *z* (plenty of elbow room).  this paper the mathematical statement of the equivalence principal follows as

... equation <Eq. 1> is also applicable to a coordinate system in which a homogeneous gravitation field is acting. In that case we have to put     = gz, where     is the gravitational potential, so that we obtain

$$ = (1 + /c^2)\, t \quad \text{<Eq. 2>} $$

The relation     = gz characterizes a uniform (i.e. homogeneous) gravitational field in Newtonian gravity.  As with any gravitational potential there is an arbitrary additive constant. This constant is chosen to be zero at the location of the clock which reads coordinate time. It was here that gravitational redshift was first postulated

There exists "clocks" that are present at locations of different gravitational potentials and whose rates can be controlled with great precision; these are the producers of spectral lines. It can be concluded from the aforesaid that the wavelength of light coming from the sun's surface, which originates from such a producer, is larger by about one part in two millionth than that of light produced by the same substance on earth.

In the footnote '#', the equivalence principle was extended to non-uniform gravitational fields

While assuming that <Eq. 2> holds for an inhomogeneous gravitational field as well.

The last section of the Jahrbuch paper is entitled "The effect of gravitation on electromagnetic phenomena." In that section a similar line of reasoning is employed to analyze an "electromagnetic process" in an accelerating frame of reference. The equations contain the factor $c(1 + gz/c^2)$, the conclusion reached

> These equations too have the same form as the corresponding equations of the nonaccelerated or gravitation-free space; however, *c* is here replaced by the value
>
> $$c(1 + gz/c^2) = c(1 + \Phi/c^2)$$
>
> From this it follows that those light rays not propagating in the *z* direction, are bent by the gravitational field…"

This was Einstein's first prediction of the gravitational deflection of light.

# (1911) On the Influence of Gravitation on the Propagation of light
## Albert Einstein

Although the Equivalence Principle first appeared in the Jahrbuch article it is sometimes taken to have originated in Einstein's 1911 paper *On the Influence of Gravitation on the Propagation of light* [14]. Here the Equivalence Principle takes the following form

> In a homogeneous gravitational field (acceleration of gravity *g*) let there be a stationary systems of co-ordinates K, oriented so that the lines of force of the gravitational field run in the negative direction of the axis of *z*. In a space free of gravitational fields let there be a second system of co-ordinates K', moving with uniform acceleration (*g*) in the positive direction of the *z* axis…
> Relatively to K, as well as relatively to K', material points which are not subjected to the action of other material points, move in keeping with the equations
>
> $$d^2x/dt^2 = 0, \quad d^2y/dt^2 = 0, \quad d^2z/dt^2 = -g$$
>
> … we arrive at a very satisfactory interpretation of this law of experience; if we assume that the system K and K' are physically exactly equivalent, that is, if we assume that we may just as well regard the system K as being a space free from gravitational fields, if we then regard K as uniformly accelerated. This assumption of exact physical equivalence makes it impossible for us to speak of the absolute acceleration of the system of reference, just as the usual theory of relativity forbids us to talk of the absolute velocity of a system; and it makes the equal falling of all bodies in a gravitational field a matter of course.

The paper concludes with the prediction of the deflection of light by the Sun. However this was not the first time such a prediction was made, nor was it correct. The gravitational deflection of light was also calculated, but not published, around 1784 by Henry Cavendish [15] . The first published calculation was almost 20 years later by Johann Georg von Soldner. Einstein's calculation in 1911 was .83 seconds of arc. Cavindish and Soldner predicted a deflection of .875 seconds of arc. As with Cavindish and Soldner, Einstein's predicted value of the deflection of light in 1911 was approximately half the correct value.

# (1916) The Foundation of the General Theory of Relativity
# Albert Einstein

In the year following the completion and publication of the general theory of relativity Einstein published the review article *The Foundation of the General Theory of Relativity* in 1916 [3]. The statement

> It will be seen from these reflections that in pursuing the general theory of relativity we shall be led to a theory of gravitation, since we are able to "produce" a gravitational field merely by changing the system of co-ordinates.

identifying gravity with non-inertial systems, wholly reflects Einstein's interpretation of gravity, the existence of which depends on the observer's frame of reference.

Facilitating the development of relativity was a landmark paper by Minkowski defining the concept of space-time [16]. Space-time is a four-dimensional "space" defined as the union of all possible events where an event is defined by a place and a time (where and when). Using the notation $x^0 = ct$, $x^1 = x$, $x^2 = y$, $x^3 = z$ an event can be written compactly as $\mathbf{x} = x^\mu = (x^0, x^1, x^2, x^3) = (ct, x, y, z)$ where *c* is the speed of light in a vacuum. If one were to plot an objects position as a function of time we would have a curve in space-time, the so-called world-line of the object. An important development, introduced by Minkowski, is the notion of the space-time interval $ds^2$. For an inertial system, in the absence of matter, the *space-time interval*, defined relative to two events separated in space-time by an infinitesimal distance, is defined as

$$ds^2 = -c^2 dt^2 + dx^2 + dy^2 + dz^2 \qquad \text{<Eq 3>}$$

where the quantity $ds^2$ is a (Lorentz) scalar, i.e. a one component object that remains unchanged under a coordinate transformation (also referred to as a tensor of rank zero. Also known as an *invariant*).

    If $ds^2 < 0$ the interval, is said to be time like. If $ds^2 > 0$ then the interval is said to be space like. If $ds^2 = 0$ then the interval is said to be light like. A worldline is space like, time like or light like according to the value $ds^2$ on the entire worldline. All particles with non-zero rest mass move on time like worldlines. If the space-time interval is time like then we can define quantity $d\tau > 0$ through the relation $ds^2 = -c^2 d\tau^2$. In this case $d\tau$ is referred to as the proper time between the two events (when the space-time interval is not infinitesimal the proper time between two events will, in general, depend on the worldline between the two events). It is the time recorded on a clock, which passes through both events on a given worldline. If we define the quantities $\eta_{\mu\nu}$ as $\eta_{00} = -1$, $\eta_{11} = \eta_{22} = \eta_{33} = 1$, and zero otherwise, then we may write the space-time interval as

$$ds^2 = \eta_{\mu\nu} dx^\mu dy^\nu \qquad \text{<Eq 4>}$$

($\mu, \nu = 0, 1, 2, 3$) where Einstein's summation convention is followed: If an index occurs twice in one term of an expression, once as a subscript and once as a superscript, it is always to be summed unless stated otherwise. The quantity $\eta_{\mu\nu}$ is known as the Minkowski metric (note: The metric is a tensor and as such it is an object that has a coordinate independent character. However a metric is only referred to as the Minkowski metric when it has the value $\eta_{\mu\nu}$). Physically, when <Eq 4.> holds in a finite region of space-time then that region is called a Minkowski space-time. If we change from the space-time system K(ct, x, y, z) to a new system K'(ct', x', y', z') then the space-time interval takes the form

$$ds^2 = g_{\mu'\nu'} dx^{\mu'} dy^{\nu'} \qquad \text{<Eq 5>}$$

where the prime represents the new coordinates, hence the invariant nature of the space-time interval. Einstein referred to theses ten space-time functions $g_{\mu\nu} = g_{\nu\mu}$ as the *fundamental tensor* or *metric tensor*. The Minkowski metric is an example of a metric tensor. Einstein identifies these quantities as the gravitational field in the following manner

The case of the ordinary theory of relativity arises out of the case here considered, if it is possible, by reason of the particular relations of the g in a finite region, to choose the system of reference in the finite region in such a way that the g assume the constant values [ $_\mu$ ]. We shall find hereafter that the choice of such co-ordinates is, in general, not possible for a finite region. ... the quantities g are to be regarded from the physical standpoint as the quantities which describe the gravitational field in relation to the chosen system of reference. For, if we now assume the special theory of relativity to apply to a certain four-dimensional region with co-ordinates properly chosen, then the g have the values [ $_\mu$ ]. A free material point then moves, relatively to this system, with uniform motion in a straight line. Then if we introduce new space-time coordinates $x^0, x^1, x^2, x^3$, by means of any substitution we choose, the g in this new system will no longer be constants but functions of space and time. At the same time the motion will present itself in the new coordinates as a curvilinear non-uniform motion, and the law of this motion will be independent of the nature of the moving particle. We shall interpret this motion as a motion under the influence of a gravitational field. We thus find the occurrence of a gravitational field connected with a space-time variability of the g . So, too, in the general case, when we are no longer able by a suitable choice of co-ordinates to apply the special theory of relativity to a finite region, we shall hold fast to the view that the g describe the gravitational field.

The geodesic equation was then derived by requiring the integral of ds be stationary between two points *P* and *P'* in space-time. This procedure results in the equation of motion, also known as the geodesic equation (actually a set of four equations)

$$d^2x^\mu/d\ ^2 + \ ^\mu\ (dx\ /d\ )(dx\ /d\ ) = 0 \qquad \text{<Eq. 6>}$$

where parameterizes the path (where is an affine parameter). The solution is a path in space-time known as a geodesic. The quantities (g $_,$ is the partial derivative of g with respect to x )

$$^\mu\ = (1/2)\ g^\mu\ (g_{\ ,\ } + g_{\ ,\ } - g_{\ ,\ }) \qquad \text{<Eq. 7>}$$

are the so-called Christoffel symbols or affine connection (more specifically the Christoffel symbols of the second kind). $g^\mu$ is the inverse matrix of $g_\mu$ , i.e. g g = . Einstein then introduces the Riemann tensor and explains its significance

$$R = \,_, - \,_, + \,_\mu{}^\mu - {}^\mu \qquad \text{<Eq. 8>}$$

> The mathematical importance of this tensor is as follows: If the continuum is of such a nature that there is a co-ordinate system with reference to which the $g_\mu$ are constants, then all the R vanish. If we choose any new system of co-ordinates in place of the original ones, the $g_\mu$ referred thereto will not be constants, but in consequence of its tensor nature, the transformed components R will still vanish in the new system. Thus the vanishing of the Riemann tensor is a necessary condition that, by appropriate choice of the system of reference, the $g_\mu$ may be constants. In our problem this corresponds to the case in which, with a suitable choice of the system of reference, the special theory of relativity holds good for a *finite* region of the continuum.

Although Einstein describes the metric tensor as the quantities which describe the gravitational field he describes it is the Christoffel symbols which are defined as the components of the field

> If the $^\mu$ vanish, then the point moves uniformly in a straight line. These quantities therefore condition the deviation of the motion from uniformity. They are the components of the gravitational field

The paper concludes with the correct values of the deflection of light by the sun by 1.7" of arc, .02" for Jupiter and a prediction of 43" of arc for the precession of Mercury.

THE GRAVITATIONAL FIELD

Consider an observer at rest with respect to a non-inertial frame of reference. Such an observer may consider himself *at rest* in this frame. Objects, which move in a uniformly as determined by an inertial observer will undergo accelerated motion as determined by the non-inertial observer. Thus acceleration is an observer dependant quantity. It is well known that the relation **F** = m**a** may be maintained *if inertial forces* are introduced. These forces are due to accelerations that result from the observer's non-inertial frame of reference. With a few notable exceptions [17,18] these forces are not considered 'real' and as such are referred to as *pseudo* or *fictitious-forces*. The gravitational force is also an inertial force since gravitational acceleration is observer dependent. This seems to be among the reasons that Einstein's interpretation of gravity has been rejected. It is the rejection of the notion that a field either exists or it doesn't. It is for this reason that Einstein is said to have shown that gravity is not a force. However the opposite is true. Einstein's intention was to show that inertial forces were 'real'. That the relation of gravity to inertia was the motivation for general relativity is expressed in an article Einstein wrote which appeared in the February 17, 1921 issue of *Nature* [19]

> Can gravitation and inertia be identical? This question leads directly to the General Theory of Relativity. Is it not possible for me to regard the earth as free from rotation, if I conceive of the centrifugal force, which acts on all bodies at rest relatively to the earth, as being a "real" gravitational field of gravitation, or part of such a field? If this idea can be carried out, then we shall have proved in very truth the identity of gravitation and inertia. For the same property which is regarded as inertia from the point of view of a system not taking part of the rotation can be interpreted as gravitation when considered with respect to a system that shares this rotation. According to Newton, this interpretation is impossible, because in Newton's theory there is no "real" field of the "Coriolis-field" type. But perhaps Newton's law of field could be replaced by another that fits in with the field which holds with respect to a "rotating" system of co-ordinates? My conviction of the identity of inertial and gravitational mass aroused within me the feeling of absolute confidence in the correctness of this interpretation."

Given that gravity and inertia were interpreted by Einstein to be one in the same phenomena, what mathematical quantity was associated with the name *gravitational field* in EGR? As we have seen above Einstein made two statements in the 1916 paper which are associated with the term *gravitational field*

> ... the quantities $g_{\mu\nu}$ are to be regarded from the physical standpoint as the quantities which describe the gravitational field in relation to the chosen system of reference.

> If the $\Gamma^\mu_{\nu\sigma}$ vanish, then the point moves uniformly in a straight line. These quantities therefore condition the deviation of the motion from uniformity. They are the components of the gravitational field.

However these statements taken together may seem rather confusing. Let us turn to Newtonian mechanics for an analogy. What mathematical quantity is given the name *gravitational field* in Newtonian gravity? Let $\Phi(\mathbf{r})$ be the Newtonian gravitational potential. Define **g** as

$$\mathbf{g}(\mathbf{r}) = -\text{grad } \Phi(\mathbf{r})$$

Then $\mathbf{g}(\mathbf{r}) = g_x(\mathbf{r})\mathbf{i} + g_y(\mathbf{r})\mathbf{j} + g_z(\mathbf{r})\mathbf{k}$ is the acceleration due to gravity at the spatial location $\mathbf{r}$, where $g_k = -(\mathbf{grad}\ \Phi)_k$, $k = x, y, z$. With these relations in mind we can establish the following analogy

$$g_\mu \Leftrightarrow \Phi \qquad \Gamma^k \Leftrightarrow g_k$$

In weak gravitational fields the equations of motion reduce to Newtonian equations as they must (let c=1 and " $\Phi,i$" is the partial derivative of $\Phi$ with respect to $x^i$)

$$\Phi = -(1/2)(g_{00} + 1)$$

$$d^2x^k/ds^2 = -\Gamma^k{}_{00} = -\Phi,_k$$

In Newtonian gravity $\Phi$ may be called the gravitational field. Most often it is the derived object $\mathbf{g}(\mathbf{r}) = -\text{grad } \Phi(\mathbf{r})$ that is called the gravitational field. Tolman would also refer to the quantities $g_\mu$ as the *gravitational potentials* [20]. Pauli referred to the totality of the $g_\mu$ values, i.e. **g**, as the *G-field* [21].

If the gravitational field is uniform in a given region of space-time then general relativity states that the field may be completely transformed away. While this is true in the special case where tidal forces are absent it is not true in general. In non-uniform gravitational fields an observer *can* tell he is in a gravitational field. Tidal gradients, i.e. space-time curvature, can neither be created nor eliminated by a change of co-ordinates. Gravitational fields *cannot*, in general, be replaced by a state of acceleration of the system in a finite region. Einstein stated this several times

> Of course we cannot replace any arbitrary gravitational field by a state of motion of the system without a gravitational field, any more than, by a transformation of relativity, we can transform all points of a medium in any kind of motion to rest [3].

> Now we might easily suppose that the existence of a gravitational field is only an *apparent* one. We might also think that, regardless of the kind of gravitational field which may be present, we could always choose another reference-body such that *no* gravitational field exists with reference to it. This is by no means true for all gravitational fields, but only for those of quite special form. It is, for instance, impossible to choose a body of reference such that, as judged from it, the gravitational field of the earth (in its entirety) vanishes [22].

Space-time regions of finite extent are, in general, not Galilean, so that a gravitational field cannot be done away with by any choice of co-ordinates in a finite region [23].

In such cases the field was later to become known as a *permanent* gravitational field [24]. Permanent gravitational fields are *not* equivalent to uniformly accelerating frames of reference.

GRAVITATION AND SPACE-TIME CURVATURE

"**Gravity** is not a foreign and physical force transmitted through space and time. It is a manifestation of the *curvature of space-time.*" That, in a nutshell, is **Einstein's theory** [25].

In a sense this view implicitly defines gravity in MGR. Since *gravitational field* is not explicitly defined we turn to the mathematical quantity associated with space-time curvature, the curvature tensor. In MGR gravity is defined implicitly through what Chandrasekhar called the *zeroth law of gravitation* which states [26].

The condition for the absence of any gravitational field is the vanishing of the (curvature tensor)

In this sense one can say that gravity is the name MGR gives to space-time curvature. Implied by this rule is the absolute, observer independent, nature of the gravitational field. Since the vanishing of the curvature tensor is not dependant on the coordinate system, the gravitational field, in MGR, either exists or it doesn't. What EGR in were called *permanent gravitational fields* i.e. non-uniform fields, are now the only gravitational fields in MGR.

At this point we can understand the source of confusion. Since a uniform gravitational field is defined as a region where the field has no tidal gradients the zeroth law of gravitation presents a pedagogical problem; Given the zeroth law of gravitation, the phrase *uniform gravitational field* is a contradiction in terms. While the zeroth law of gravitation represents the interpretation of gravity in MGR it does not reflect Einstein's interpretation. Nothing in either EGR nor MGR prohibits the curvature tensor from vanishing when the metric $g$ is not constant. I.e. $g_{\mu\nu}$ does not imply $R_{\alpha\beta\mu\nu} = 0$. This is exactly the case for the uniform gravitational field.

Although in EGR the field is described by the metric, physically it is motion and distances that are measured. The quantities $\Gamma^\mu_{\alpha\beta}$, determine the deviation of motion from straight line. Einstein protested when Max von Laue stated that it was the Riemann tensor that was the only representation of the gravitational field [27].

> ... what characterizes the existence of a gravitational field from the empirical standpoint is the non-vanishing of the components of the affine connection], not the vanishing of the [components of the Riemann tensor]. If one does not think in such intuitive (anschaulich) ways, one cannot grasp why something like curvature should have anything at all to do with gravitation. In any case, no rational person would have hit upon anything otherwise. The key to the understanding of the equality of gravitational mass and inertial mass would have been missing.

In light of the above, according to EGR, space-time curvature is a necessary, but not sufficient, condition for the existence of gravity. It would be more accurate, in EGR, to say that space-time curvature is a manifestation of gravitation since $R_{\alpha\beta\mu\nu} \neq 0$ (space-time curvature) implies $g_{\mu\nu}$ everywhere (gravitation) but $g_{\mu\nu}$ (gravitation) does not imply $R_{\alpha\beta\mu\nu} \neq 0$ (space-time curvature).

It should also be noted that there are various terms used to describe gravity which seem quite similar. Such terms a *curvature* or *warp* are often used to mean something other than space-time curvature. The terms time warp and time curvature have been used to refer to gravitational redshift (i.e. gravitational time dilation) [28,29]. The term curvature has also been used to refer to gravitational acceleration or to the curved path an object may take [30]. These should not be confused with space-time curvature since they refer to different quantities.

NON-PERMANENT GRAVITATIONAL FIELDS

In order to shed some light on the difference between EGR and MRG it would be instructive to give a few examples of gravitational fields that have zero space-time curvature. In this section we consider three such examples.

A very important gravitational field with zero space-time curvature in general relativity is the uniform gravitational field. The gravitational acceleration of an object ini such a field will be calculated here since there will be occasion to use it below. The metric for a uniform gravitational field is [31,32,33,34].

<Eq. 9>  $ds^2 = -(1+gz/c^2)^2 d(ct)^2 + dx^2 + dy^2 + dz^2$

where $-c^2/g < z < c^2/g$. The components of the metric are $g_{00} = -(1+gz/c^2)^2$, $g_{11} = g_{22} = g_{33} = 1$. For a particle with non-zero rest mass the proper acceleration $d^2x^\mu/d\tau^2$ can be obtain by parameterizing the geodesic equation with the proper time, i.e. $\tau =$

<Eq. 10>  $d^2x^\mu/d\tau^2 + \Gamma^\mu_{\alpha\beta}(dx^\alpha/d\tau)(dx^\beta/d\tau) = 0$

To obtain $d^2z/d\tau^2 = d^2x^3/d\tau^2$ we set $\mu = 3$ into <Eq. 10> and then substitute into <Eq. 7>.

$\Gamma^3_{\alpha\beta} = (1/2) g^{33}(g_{3\alpha,\beta} + g_{3\beta,\alpha} - g_{\alpha\beta,3}) = -(1/2) g^{33} g_{\alpha\beta,3}$

since $g_{3\alpha,\beta} = g_{3\beta,\alpha} = 0$. Since the only non-vanishing $g_{\alpha\beta,3}$ is $g_{00,3} = -2(g/c^2)(1+gz/c^2)$ only $\Gamma^3_{00}$ remains. Since $g_{\mu\nu}$ is a diagonal, $g^{\mu\mu} = (g_{\mu\mu})^{-1}$, $g^{33} = 1$

<Eq. 11>  $\Gamma^3_{00} = (g/c^2)(1 + gz/c^2)$

Substituting <Eq. 11> into <Eq. 10> yields

<Eq. 12>  $d^2z/d\tau^2 = -\Gamma^3_{00}(dt/d\tau)^2 = -(g/c^2)(1 + gz/c^2)(dt/d\tau)^2$

$(dt/d\tau)^2$ can be obtained from <Eq. 9> by setting $-c^2 d\tau^2 = ds^2$ and solving for $(dt/d\tau)^2$ giving

<Eq. 11>  $(dt/d\tau)^2 = [(1+gz/c^2)^2 - v^2/c^2]^{-1}$

where $v^2 = v_x^2 + v_y^2 + v_z^2$. Substituting <Eq. 11> into <Eq. 12> yields

<Eq. 13>  $d^2z/d\tau^2 = -(g/c^2)(1 + gz/c^2)/[(1+gz/c^2)^2 - v^2/c^2]$

This expression gives the local acceleration of an object whose velocity is $v$. Observers at different positions z in the field system will *not* measure the same local value of acceleration. Not only will the object's acceleration depend of position but also on velocity. This is contrary to what one would normally assume for a uniform gravitational field. In fact some workers *define* a uniform gravitational field according to the relation (assuming $v = 0$) [35,36].

<Eq. 14>    $d^2z/d\tau^2 = -g$

However both <Eq. 13> and <Eq. 14> cannot both true.

Another example of a gravitational field with zero space-time curvature is that of a vacuum domain wall [37,38]. A domain wall is two a dimensional structure characterized by a fixed energy per unit area $\sigma$. A surface tension equal to $\sigma$ (in appropriate units) is required in order for $\sigma$ to remain constant. The field acts as if had a negative pressure in the two directions within the wall. If the wall lies in the xy-plane then the space-time interval may be written as

<Eq. 15>  $ds^2 = -(1 - g|z|/c^2)^2 d(ct)^2 + (1 - g|z|/c^2)^2 e^{-2\kappa ct}(dx^2 + dy^2) + dz^2$

where $g = 2\pi G\sigma$. This coordinate system is that of an observer at $z = 0$ at rest relative to the wall. An observer will see a test particle undergo a hyperbolic motion with proper acceleration $g$ away from the wall acceleration. The wall thus acts like it has a negative active gravitational mass. The curvature tensor for this gravitational field is $R_{\mu\nu} = 0$ everywhere expect inside the wall itself.

Another space-time with zero curvature is that of a straight cosmic string. The composition of such a string is reminiscent of a domain wall [37,38]. However a string is a one dimensional object with constant energy per unit length $\mu$. The magnitude of the string's tension is also equal to $\mu$ (in appropriate units). The negative active gravitational mass associated with the tension balances the positive gravitational mass. The result is the absence of gravitational acceleration. Such a string exerts no gravitational force on objects in the surrounding space. The space-time interval for this string is

<Eq. 16>    $ds^2 = -d(ct)^2 + dr^2 + (1 - 8G\mu)r^2 d\phi^2 + dz^2$

As mentioned above the gravitational force on objects in the surrounding space is zero. However string the alters does alter the surrounding space making it non-Euclidean. The factor $(1 - 8G\mu)$ implies that a circle at fixed r, z, t has a reduced circumference given by

<Eq. 17>    $C = 2\pi(1 - 4G\mu)r$

This indicates that an angle of $\Phi = 8\pi G\mu$ is missing from the space surrounding the string (See Figure 1). The geometry is non-Euclidean. Vilenken has noted that the geometry is conical (See Figure 2). Light passing by the string is deflected by an angle $\Delta = 4\pi G\mu$. An object behind the string will then appear as a double object to an observer on the other side of the string. The string thus acts as a gravitational lens. Gravitational lensing is usually discussed in terms of curved space-time.

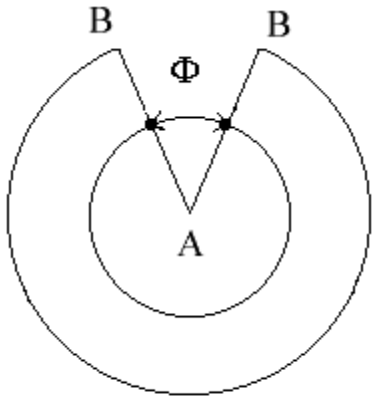
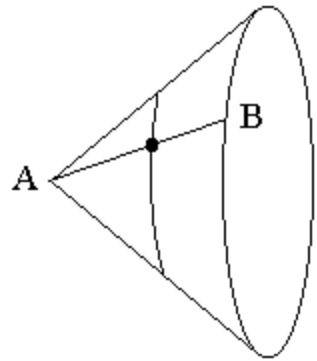

Figure 1                    Figure 2

In the case of the cosmic string the lensing is a result of the non-Euclidean nature of the surrounding space [39]. $R_\mu = 0$ everywhere except at $r = 0$, where the curvature is undefined, even though the metric is well defined everywhere. Only on the string itself, is $R_\mu$ undefined. This can be seen by considering the geometry of a cone. The curvature of the surface of a cone is well defined except at the apex.

EXAMPLES FROM THE LITERATURE

In this section we review various instances where the change of interpretation may have caused some difficulty. The main problem with the change in definition of gravity is that one expects to find curvature where none should be found. As stated above the equivalence principle, the basis of general relativity, becomes a contradiction in terms in modern relativity.

When asserting that gravity is a curvature in space-time one might conclude that any "true" gravitational field has tidal forces. But what is meant by a "true" gravitational field? We have seen above that gravitational effects can be observed in the absence of tidal gradients. However a uniform gravitational field has no tidal gradients and can be completely transformed away. A freely falling observer will not be able to detect the presence of the field. His frame is an inertial one. Conversely, when the observer changes his system of coordinates he can produce a uniform gravitational field. In both cases the curvature tensor vanishes. If one uses the modern definition of gravity then one might expect the curvature tensor to be non-zero. This can be very confusing and is reflected by a comment made by Synge [40].

> ... I have never been able to understand this principle…Does it mean that the effects of a gravitational field are indistinguishable from the effects of an observer's acceleration? If so, it is false. In Einstein's theory, either there is a gravitational field or there is none, according as the Riemann tensor does not or does vanish. This is an absolute property; it has nothing to do with any observers world line … The Principle of Equivalence performed the essential office of midwife at the birth of general relativity, but, as Einstein remarked, the infant would never have gone beyond its long clothes had it not been for Minkowski's concept [of space-time geometry]. I suggest that the midwife be buried with appropriate honours and the facts of absolute space-time faced.

A similar example is that of Ray [41].

> It is very important to notice that in a freely falling frame we have not transformed away the gravitational field since the Riemann tensor (gravitation    Riemann tensor) will not vanish and we will still measure relative acceleration ….. The first thing to note about the 1911 version of the principle of equivalence is that what in 1911 is called a uniform gravitational field ends up in general relativity not to be a gravitational field at all – The Riemann tensor is here identically zero. Real gravitational fields are not uniform since they must fall off as once recedes from gravitating matter.

Here, too, the presence of gravity is associated with curvature. An observer falling in a non-uniform gravitational field will be able to detect curvature since tidal forces are not transformed away. However the equivalence principle applies to uniform fields and was not meant to apply otherwise (it only applies locally). Contrary to Ray's assertion a uniform gravitational field can be created from a finite distribution of mass. For example; Consider a spherical body of uniform mass density. If a spherical cavity is excavated such that the center of the cavity is not concentric with the sphere then the field inside will be highly uniform, i.e. there will be virtually no tidal forces present.

Desloge attempted to prove the equivalence principle wrong with the following argument [36],

> Gravity is a manifestation of the local curvature of space-time and the local curvature of space-time is a measurable quantity. Hence, observers, regardless of their motion, can always determine whether or not they are in a gravitational field by determining whether or not space-time is curved in their neighborhood. .... In a uniform gravitational field the quantity $g(x)$, the initial local acceleration of a particle released from rest at x, must have the same value at all points in the field. If this were not true, the field would not be uniform.

The metric for a uniform gravitational field then derived based on these assumptions. The metric which results is then used to calculate the curvature tensor which does not vanish. However Desloge expects curvature tensor to vanish

> It follows that in a uniform gravitational field space-time is as expected curved.

What happened? Consider a uniformly accelerating frame of reference. From what we have seen above, for an object at rest at $z = 0$, the initial acceleration of an object falling from rest is $-g$. But, as seen in the previous section, an object initially fixed at z and allowed to fall, the initial acceleration in a local frame of reference will be $-g/(1+gz/c^2)$. The initial acceleration of a freely falling object decreases as one moves to points high in the field. Deloge's assertion

> *the initial local acceleration of a particle released from rest at x, must have the same value at all points in the field.*

is, in reality, an assumption on the nature of a falling object in a uniform gravitational field. The origin of the problem here is that Deloge is defining a gravitational field according to the definition that the local acceleration due to gravity is the same at all places in the field. This is reminiscent of the uniform gravitational field in Newtonian gravity, that the gravitational acceleration is a constant independent of position. However in Newtonian gravity this is equivalent to saying that there are no tidal gradients in the field. This definition is inherent in the way Einstein imagined falling off a roof. In free fall objects at rest relative to him would remain at rest. So in the relativistic case it's quite natural to take as the definition of the uniform gravitational field as a field with no tidal forces and thus by definition the Riemann tensor is zero – no space-time curvature. This definition is consistent with the Equivalence Principle. Thus Desloge has simply redefined "uniform" in a manner which resulted in a non-vanishing Riemann tensor. Since this was expected the end result of the presence of tidal forces was not questioned and thus the starting definition not questioned.

Gravitational redshift is often interpreted as the result of space-time curvature [42, 43,44,45]. The argument used to arrive at this conclusion is based on that proposed by Schild [46] or some variation [47]. The validity of this interpretation has previously been discussed elsewhere in more detail [48]. The presence of curvature is neither a necessary nor sufficient condition for the detection of gravitational redshift. If a light source is located at $\mathbf{r}_1$ and a light detector at $\mathbf{r}_2$ and they remain fixed at thes positions then the necessary and sufficient condition for gravitational redshift is $g_{00}(\mathbf{r}_1) \ne g_{00}(\mathbf{r}_2)$. This in no way requires $R_{\mu\nu} \ne 0$. In fact if one performs this experiment in an enclosed room (in which there is gravitational field present or in a room which is accelerating uniformly) with no knowledge of the environment outside (We don't assume that the room is at rest on the surface of the Earth) then it would be impossible to tell if there were curvature present. Only if the light emitter or detector is moved to various locations and seperate readings taken can one determine whether $R_{\mu\nu} \ne 0$ or not. However this was not the case in the classic gravitational redshift experiments performed by Pound and [49]. In fact the first calculation of redshift ever given was that by Einstein in 1907 for an accelerating frame of reference and hence for a uniform gravitational field. It was that relationship which was tested by Pound and Rebka.

CONCLUSION

Einstein did not interpret gravity as a curvature of space-time, rather that space-time curvature is a manifestation of gravity. This is the difference between EGR and MGR, one that may certainly be confusing. This change in view from Einstein's to the modern view perhaps reflects the desire of attributing "real" quantities as having an absolute existence independent of the observer i.e. "Either something exists or it doesn't." However Einstein showed us a very different way of looking at reality. In fact Einstein was suspicious of the very idea of "real" [50].

> It appears to me that "real" is an empty meaningless category (drawer) whose immense importance lies only in that I place certain things inside it and not certain others.

If one wishes to understand the various problems which have arisen one should a complete understanding of the terms used, how they are used and the possible misunderstandings that could arise from their use. As Einstein phrased it [51]

> …, a hypothesis is a statement whose *truth* is temporarily assumed, *whose meaning is beyond all doubt*

Nature fails to behave in a manner that allows definitions that are acceptable to all. There will at times be disagreement among individuals with opposing views. Such disagreements will, for the most part, be over the appropriate choice of words to use and is not relevant to experimental predictions except in that they influence and direct the creative processes from whence those predictions are made. Nature has Her own way of doing things regardless of what physicists believe. Einstein was not unaware of such disagreement [52].

> But because the Lord had noticed already before the development of theoretical physics that He cannot do justice to the opinions of man, He simply does as *He* sees fit.

ACKNOWLEDGEMENTS

The author would like to thank the following people for their advice and constructive criticism. In no particular order: Edwin F. Taylor, Robert W. Brehme, Hans C. Ohanian, John Stachel, and A.P. French. Any errors which may remain are the sole responsibility of the author.